\documentclass{acm_proc_article-sp}
\usepackage{multirow}

\begin{document}

\title{Feasibility Evaluation of 6LoWPAN over Bluetooth Low Energy}

\numberofauthors{2}
\author{
\alignauthor
Varat Chawathaworncharoen% \\
       \titlenote{This work was done during an internship at AIST, Japan.},\\
Vasaka Visoottiviseth\\
       \affaddr{Mahidol University}\\
       \affaddr{Nakorn pathom, Bangkok 73170, Thailand}\\
       \email{ultimaxx7@gmail.com, vasaka.vis@mahidol.ac.th}
\alignauthor
Ryousei Takano\\
       \affaddr{National Institute of Advanced Industrial Science and Technology~(AIST)}\\
       \affaddr{Tsukuba, Ibaraki 305-8568, Japan}\\
       \email{takano-ryousei@aist.go.jp}
}

\maketitle
\begin{abstract}
IPv6 over Low power Wireless Personal Area Network~(6LoWPAN) is an emerging
technology to enable ubiquitous IoT services.  However, there are very few
studies of the performance evaluation on real hardware environments.
This paper demonstrates the feasibility of 6LoWPAN through conducting
a preliminary performance evaluation of a commodity hardware
environment, including Bluetooth Low Energy~(BLE) network, Raspberry Pi,
and a laptop PC.
Our experimental results show that the power consumption of 6LoWPAN over BLE
is one-tenth lower than that of IP over WiFi; the performance significantly
depends on the distance between devices and the message size; and
the communication completely stops when bursty traffic transfers.
This observation provides our optimistic conclusions on the feasibility of
6LoWPAN although the maturity of implementations is a remaining issue.
\end{abstract}

\if0
% A category with the (minimum) three required fields
%A category including the fourth, optional field follows...

\fi


%==============================================================================
\section{Introduction}

The Internet of Things~(IoT) is an IT system based on a network of smart
objects embedded with a sensor, software, and connectivity to exchange data with
service providers.
In the vision of Trillion Sensors Universe~\cite{Bryzek2013.roadmap}, for
example, a tremendous amount of smart objects will be deployed everywhere on
the earth, and each of them has connectivity to the Internet, either directly
or indirectly through gateways.
To provide such a ubiquitous connectivity with
several restrictions to power, memory space, network bandwidth, and processing
resources~\cite{rfc7228}, IPv6 over Low-power Wireless Personal Area
Network~(6LoWPAN)~\cite{rfc6282}, which is standardizing by IETF, is a
promising technology.

In terms of the link layer, several low-power wireless technologies, including
ZigBee/IEEE 802.15.4, Bluetooth Low Energy, and Wi-SUN, have been developed for
supporting IoT applications and services.
Bluetooth Low Energy~(BLE)~\cite{Bluetooth} or Bluetooth Smart aims at
enabling low-cost sensors to exchange data for short distance,
and it has a wide range of applications, including smart watches,
home electronics, location-based services such as Apple's iBeacon and
Google's Eddystone.
The Bluetooth specification version 4.1 or newer is required for IPv6 over BLE
links~\cite{6loble}.
Although the standardization process is ongoing, some operating systems have
already supported it in advance.

In this paper, we demonstrate the feasibility of 6LoWPAN over BLE through
conducting experiments on a commodity hardware environment.
Our contribution is a preliminary performance evaluation of
6LowPAN over BLE,
including the power consumption comparing with both wired and wireless
Ethernet technologies, the impact of the distance between devices and the
message size on the throughput, and the application performance based on MQTT.

The rest of the paper is organized into the following sections:
Section~\ref{sec:protocol} presents the overview of a network protocol stack
supporting IoT services,
Section~\ref{sec:experiment} shows our experimental results in a commodity
software and hardware environment,
Section~\ref{sec:usecase} demonstrates a simple MQTT application as a use case of IoT services,
and finally Section~\ref{sec:conclusion} summarizes the paper and briefly
mentions future work.

%==============================================================================
\section{IoT Protocols}\label{sec:protocol}

\begin{figure}
\includegraphics[width=.4\textwidth]{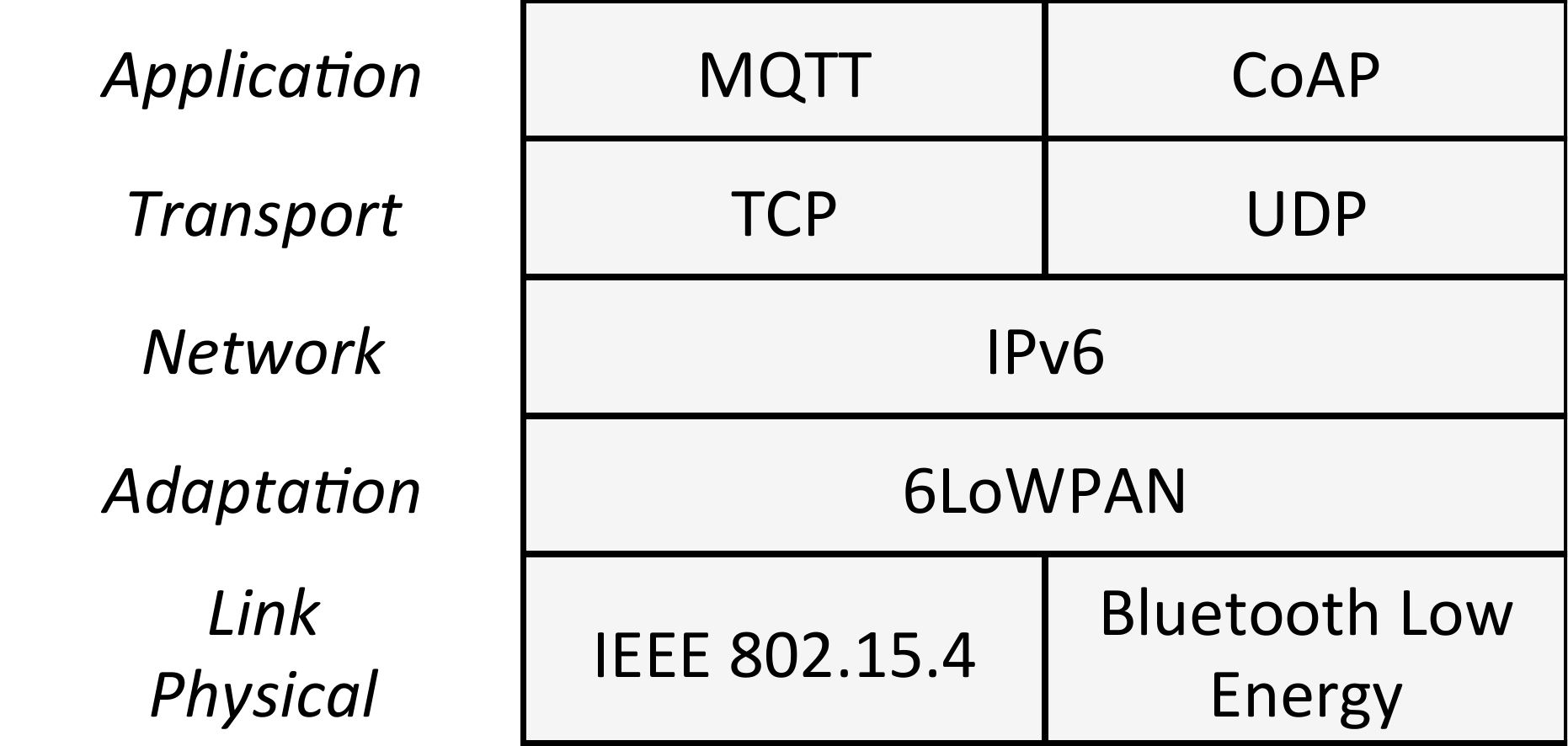}
\caption{6LoWPAN protocol stack}
\label{fig:6lowpan}
\end{figure}

Figure~\ref{fig:6lowpan} shows a typical 6LoWPAN protocol stack from the
physical layer to the application layer.
Though 6LoWPAN is originally designed for IEEE 802.15.4-based networks
~\cite{rfc6282}, currently 6LoWPAN over BLE is under the process of
standardization at IETF~\cite{6loble}.
The physical bit rate of BLE is up to 1~Mbps, and the effective throughput is
about one-third of it.
BLE has two roles of devices: a master and a slave.
A slave device broadcasts advertise messages until a master detects it.
After the link layer connection establishment, 6LoWPAN initialized the network
interface, and IPv6 communication between them is ready to start.

\begin{table*}
\centering
\caption{Comparison among application protocols}
\label{tbl:protocols}
\begin{tabular}{p{3cm}|p{4cm}p{4cm}p{4cm}} \hline
 & MQTT & MQTT-SN & CoAP \\
\hline
Abstract       & PubSub & PubSub & REST \\
Architecture       & Broker & Broker & client-server \\
Transport protocol & TCP & TCP/UDP & UDP \\
Encoding           & Binary & Binary & Binary \\
IP Multicast       & No & No & Yes \\
QoS         & 3 levels (At most/At least/Exactly once delivery) & 3 levels (At most/At least/Exactly once delivery) & Confirmable messages \\
Security           & Simple password authentication, SSL/TLS & No & DTLS \\
Standardization    & OASIS~\cite{MQTT} & Not yet (IBM~\cite{MQTT-SN}) & IETF~\cite{CoAP} \\
Implementation     & Mosquitto~(broker)~\cite{Mosquitto},
Paho~(client)~\cite{Paho} & gateway and client~\cite{MQTT-SN.impl} & Ponte~(server), libcoap \\
\hline\end{tabular}
\end{table*}
% RT: can you show the source of NAT problem.

6LoWPAN is an adaptation layer protocol in the middle of a link layer and
a network layer, and it allows the transmission of IPv6 datagrams over IEEE
802.15.4 or BLE networks.  Precisely speaking from the viewpoint of the BLE network,
6LoWPAN works on top of the Logical Link Control and Adaptation Protocol
~(L2CAP) layer.
IPv6 requires to transmit datagrams of 1280 bytes or larger, and the minimum
header size is 40 bytes.  However, the physical packet size of BLE is up to 47
bytes.
6LoWPAN fills the gap by employing several features including header
compression and fragmentation.
Several header compression schemes are defined, and the proper scheme is used
for each IPv6 address type such as a link-local address and a unique local
address.
At network interface initialization, a link-local address based on the 48-bit
BLE address is automatically assigned to the device.  In this case, IPv6
header can be compressed to only 2 bytes.  However, applications cannot use a
link-local address, and a non-link-local address have to be assigned to the
network interface.  In this case, IPv6 header can be compressed to 12 or 20
bytes. % RT (FIXME)

Application protocols layer on top of transport protocols.
Table~\ref{tbl:protocols} briefly compares three popular application protocols
suitable for communication in machine-to-machine~(M2M) and IoT:
MQTT~\cite{MQTT}, MQTT-SN~\cite{MQTT-SN}, and CoAP~\cite{CoAP}.
Message Queuing Telemetry Transport~(MQTT)~\cite{MQTT} is a publish/subscribe
messaging transport protocol.
An MQTT system contains three roles: broker, publisher, and subscriber.
A broker is a medium for message exchanging among clients.  A publisher
transfers the message that be referenced by topic.  A subscriber waits
for messages that related to the topic.  If the topic that a subscriber attend
was changed, the message is distributed to the subscribers.
MQTT also provides three-level QoS for delivering messages between clients and
brokers: ``at most once'', ``at least once'', and ``exactly once''.
MQTT for Sensor Network (MQTT-SN)~\cite{MQTT-SN} is a lightweight variant of
MQTT for low bandwidth and high failure networks, and devices with significant
resource constraints.  It does not require TCP and security features.
Usually, MQTT-SN gateways transfer MQTT-SN messages to an MQTT broker.
Constrained Application Protocol~(CoAP)~\cite{CoAP} is an HTTP-like transport
protocol based on the Representational State Transfer~(REST) model.
Servers make resources available under a URL, and clients access these
resources using methods such as GET, PUT, POST, and DELETE.
Unlike HTTP, CoAP is based on UDP and the encoding is binary form.

The rest of the paper shows MQTT over 6LoWPAN because we can use several
well-designed
and open source MQTT implementations as shown in Table~\ref{tbl:protocols}.

%==============================================================================
\section{Experiment}\label{sec:experiment}
To demonstrate the feasibility of 6LoWPAN over BLE, we have conducted two
experiments.  This section shows the performance evaluation including
the power consumption and the application-level performance using MQTT
while the next section demonstrates a use case of MQTT over 6LoWPAN.

%------------------------------------------------------------------------------
\subsection{Experimental setting}
We have built a minimum 6LoWPAN environment which consists of ThinkPad X230t
~(X230) and Raspberry Pi model B+~(RasPi).
Table~\ref{tbl:spec} summarizes the specifications.
Linux operating system is running on both X230 and RasPi, where the kernel 3.18
and later have already supported 6LoWPAN over BLE.

\begin{table*}
\centering
\caption{Hardware and Software Specifications}
\label{tbl:spec}
\begin{tabular}{l|l|l} \hline
& ThinkPad X230t & Raspberry Pi model B+\\ \hline
CPU & Intel Core i7-3520M@2.90~GHz & Broadcom BCM2835~(ARM1176JZF-S)@700~MHz\\
Memory & 16~GB & 512~MB\\
Disk & SSD 256~GB & microSD Card 32~GB\\
Ethernet & Intel 82579LM & Microchip LAN9512\\
BLE & \multicolumn{2}{l}{I-O DATA USB-BT40LE} \\
WiFi & \multicolumn{2}{l}{I-O DATA WN-G150UMK} \\
\hline
OS & Ubuntu 14.04.2 LTS 64bit & Raspbian GNU/Linux 7\\
Kernel & 3.18.0-031800-generic & 3.18.14+ \\
\hline\end{tabular}
\end{table*}

To evaluate the performance of MQTT on a 6LoWPAN environment, MQTT version
3.1 compliant implementations were used as follows.
Mosquitto~\cite{Mosquitto} is an open source broker implementation written in
the C language.  We used the version 3.1.
Paho~\cite{Paho} is an open source client library and supports multiple
programming languages such as C, Java, and Python.  We used version 1.1
and, both our benchmark and application programs were written in C.

We used unique local IPv6 addresses in MQTT experiments.
After establishing 6LoWPAN connection, a link-local IPv6 address is
automatically assigned to the BLE device as described in Section~\ref{sec:protocol}.
We also assigned a unique local IPv6 address to each device because Paho C
library cannot use a link-local IPv6 address to transfer messages.

%------------------------------------------------------------------------------
\subsection{Benchmark}

\subsubsection{Power Consumption}\label{sec:power}
We measured the power consumption of RasPi on several conditions, that is,
the combination of three network devices(wired Ethernet, WiFi, and BLE) and
three workloads (idle, ping, and iperf benchmark program).
The distance between devices is 0 meter.
RasPi is supplied 5V power from USB cabling.
To measure the power consumption, we observed current on a USB power line by
Sanwa PC700 multimeter.

Table~\ref{tbl:power} shows that the absolute power consumption of each case
and the increase over the baseline, that is idle without any network
devices.
The power consumption of 6LoWPAN over BLE is one-tenth lower than that of IP
over WiFi.  BLE takes lowest power consumption among three devices, and the
increase over the baseline is only 0.01 mA.  On the other hand, WiFi takes the
highest power consumption.
In terms of the transferred per Joule, however, BLE is not efficient because
the throughput is quite small comparing with the theoretical link bandwidth,
which is 1 Mbps.  We consider the implementation may not be mature enough.

\begin{table}
\centering
\caption{Comparison of power consumption among a combination of network
devices and workloads [mA]}
\label{tbl:power}
\begin{tabular}{ll|rr} \hline
device & workload & observed & diff. \\
\hline
- & idle & 0.20 & - \\
\hline
\multirow{3}{*}{wired Ethernet} & idle & 0.24 & 0.04\\
 & ping & 0.24 & 0.04\\
 & iperf~(58.3~Mbps) & 0.27 & 0.07 \\
 % bit / joule = 43,185 Kb/Joule
\hline
\multirow{3}{*}{WiFi} & idle & 0.29 & 0.09 \\
 & ping & 0.30 & 0.10 \\
 & iperf (32.2~Kbps) & 0.32 & 0.12 \\
 % bit / joule = 20.125 Kb/Joule
\hline
\multirow{3}{*}{BLE} & idle & 0.21 & 0.01 \\
 & ping & 0.21 & 0.01 \\
 & iperf (5.84~Kbps) & 0.21 & 0.01 \\
 % bit /joule = 5.5619047619 Kb/Joule
\hline\end{tabular}
\end{table}

\subsubsection{Distance between devices}

We measured the impact of the distance between 6LoWPAN devices on the
application-level performance, that is, how many MQTT messages can be
published from a client to the broker for one second, where the distance is
varied from 0 to 20 meters.
Each benchmark trial takes 10 seconds.
This experiment was conducted in a corridor without any blind spots and
devices were put on the floor.

\begin{figure}
\centering
\includegraphics[width=.45\textwidth]{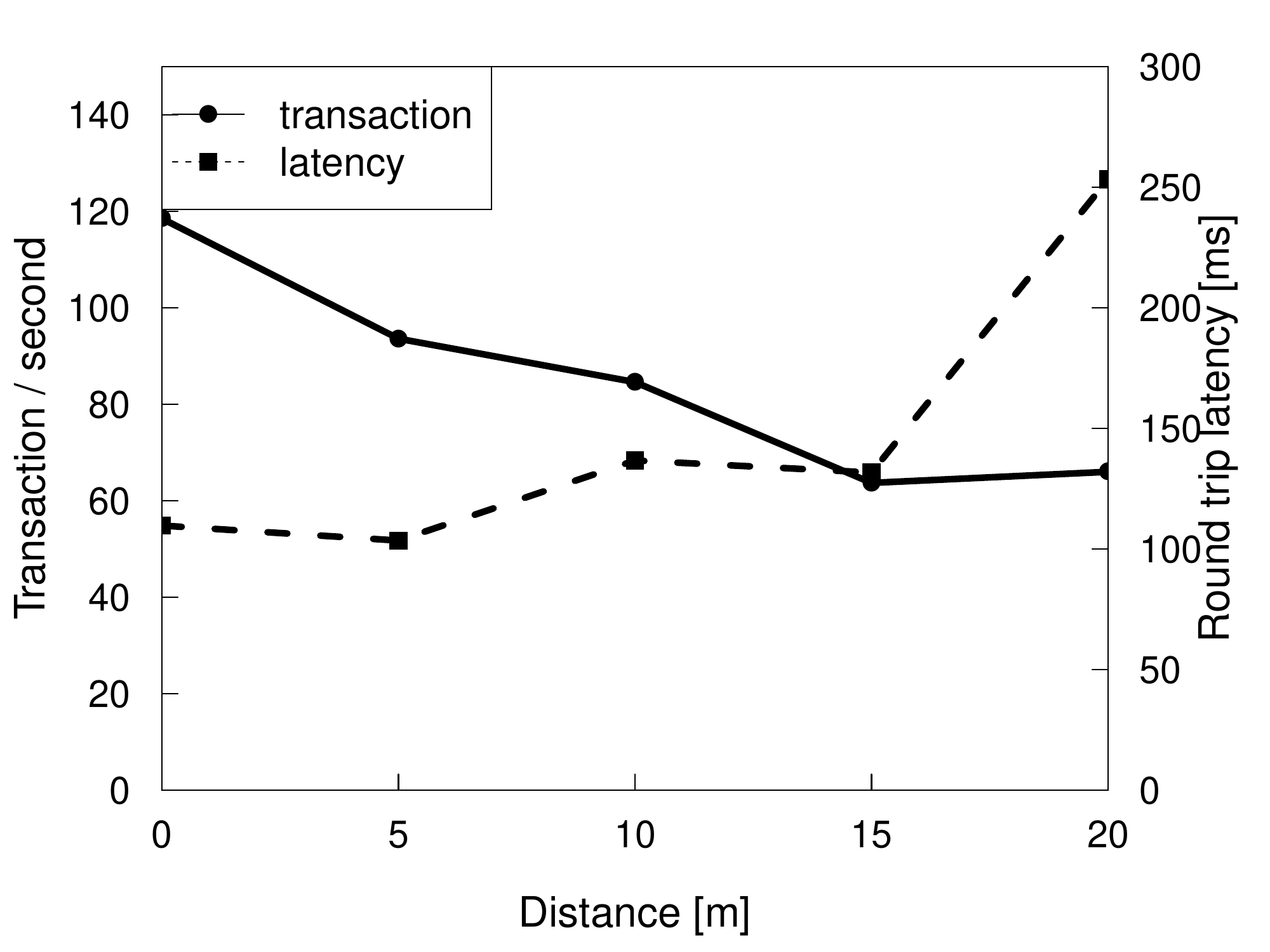}
\caption{Latency and MQTT throughput while varying the distance between devices}
\label{fig:distance}
\end{figure}

Figure~\ref{fig:distance} shows results of the round trip latency and the MQTT
throughput.  The MQTT throughput linearly decreases as the distance increases.
Finally, the communication failed where the distance is over 20 meters.
It is because that MQTT runs on top of TCP and TCP performance degrades as
the round trip time increases.
We observed a few TCP retransmissions during this experiment, where the number
of TCP retransmissions does not depend on the distance.
Note that we obtained each result separately, and the latency fluctuated;
therefore the correlation between them is not so evident from Figure~\ref{fig:distance}.

% RT: I'd like to measure it again and confirm it is reproducible.
\if0
\begin{table}
\centering
\caption{Latency and MQTT throughput while varying the distance between devices}
\label{tbl:distance}
\begin{tabular}{r|rr} \hline
Distance [m] & ping [ms] & Transaction/sec \\
\hline
0  & 109.7 & 118.5 \\
5  & 103.5 & 93.6 \\
10 & 136.7 & 84.6 \\
15 & 131.7 & 63.7 \\
20 & 235.3 & 66.0 \\
\hline\end{tabular}
\end{table}
\fi

\subsubsection{Message size}
We also measured the impact of the message size on the MQTT throughput where
the message size is varied from 1 to 256 bytes.
The distance between devices is 0 meter.
Each benchmark trial ran in the period of 10 seconds.

\begin{figure}
\centering
\includegraphics[width=.45\textwidth]{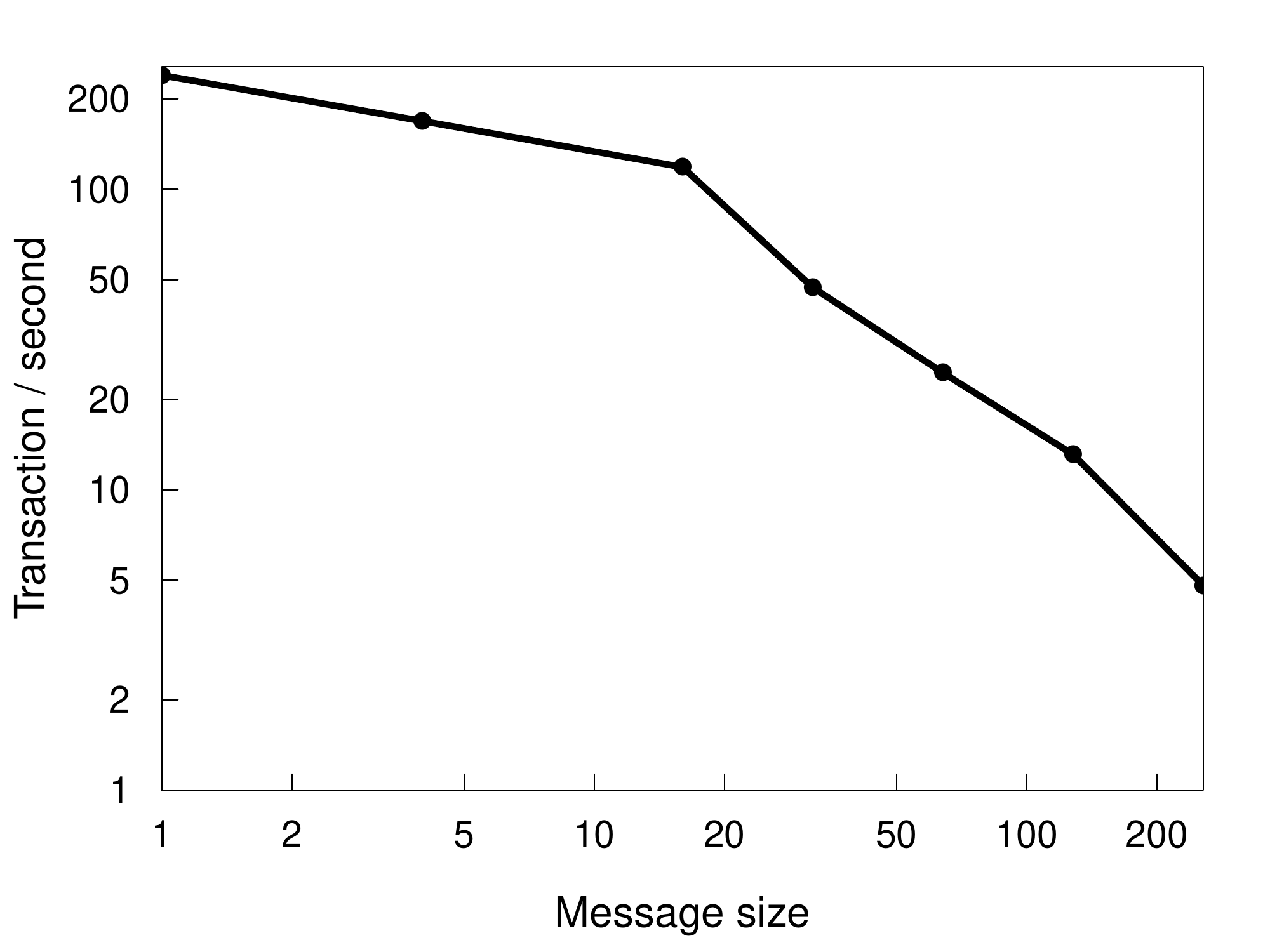}
\caption{MQTT throughput while varying the message size~(log-log plot)}
\label{fig:msgsize}
\end{figure}

Figure~\ref{fig:msgsize} presents the relationship between the throughput and
the message size.
The message size is a significant factor for the performance.  The throughput
decreases in proportion to the message size, and it suddenly drops between
16-byte and 32-byte messages.  It can be caused by fragmentation, where
the message is fragmented into multiple BLE packets.
This result leads that application programmers should keep the message
size as small as possible to get better performance.
% RT: does segmentation (message is divided into several 6LoWPAN packets with the MTU size) affect the throughput?  It seems there is a performance gap between 16-byte message and 32-byte message.
% RT: how long is the payload size of BLE?

\if0
\begin{table}
\centering
\caption{MQTT throughput while varying the message size}
\label{tbl:msgsize}
\begin{tabular}{l|l} \hline
Message size [byte] & Transaction/sec \\
\hline
1   & 239.7 \\
4   & 168.7 \\
16  & 118.5 \\
32  & 47.1 \\
64  & 24.5 \\
128 & 13.1 \\
256 & 4.8 \\
\hline\end{tabular}
\end{table}
\fi

%==============================================================================
\section{Use Case: USB-powered Light Application}\label{sec:usecase}
We have implemented a simple MQTT application as a use case of IoT services.
This application allows the users to control the USB-powered light from remote
computers, and it is quite simple but enough for demonstrating the
usability and the functionality of MQTT.

% design
This application consists of two clients: {\it device} and {\it controller}.
A device client is running on a computer that has a target device, e.g., USB
powered-light, and controls it. On the other hand, a controller client
controls the target device and get the status remotely.
These clients exchange the MQTT message via a broker, and they act as
sometimes subscribers and at other time publishers.
We define two topics: ``light/control'' and ``light/status'', as shown in
Table~\ref{tbl:topic}.  The former is used to turn the power on and off; the
latter is used to get the power status.  Besides, to keep the latest power
status on the broker, we set messages for topic ``light/status'' to be
retained.
% RT: you should add the definitions of topic and message in Section 2.
% RT: I don't think device and controller are good naming. The terms come from https://bitbucket.org/ryousei/mu2015/wiki/firstapp. Do you have another idea?

% implementation
This application is written in Python with the Paho Python library.
To control the power supply of each USB port, we use {\it hub-ctrl}
~\cite{hub-ctrl} on a controller client.
In this experiment, device and controller clients ran on RasPi and X230,
respectively; the broker ran on X230.

\begin{table}
\centering
\caption{Topics of USB-powered light application}
\label{tbl:topic}
\begin{tabular}{l|p{6cm}} \hline
Topic & Description \\ \hline
light/control & This topic is used to turn the light on and off.
The message should contain ``on'' or ``off'', otherwise the client returns an
error. \\
light/status & This topic is used to get the status of the light.
It returns ``on'' or ``off''.\\
\hline\end{tabular}
\end{table}

% issue
We found a critical problem that hub-ctrl turns off all of USB ports on
RasPi.  Therefore, the ``light/control off'' message turns off not only the
light but also the BLE USB dongle.
By using GPIO instead of hub-cntl, we can control AC power supply to
external electronics products like this application.

%==============================================================================
\section{Conclusion and Future Work}\label{sec:conclusion}

6LoWPAN is a promising technology in the IoT era.
To demonstrate the feasibility, we have conducted a preliminary
performance evaluation of a commodity hardware environment,
including Bluetooth Low Energy~(BLE) network, Raspberry Pi, and a laptop PC.
Our experimental results show that the power consumption of 6LoWPAN over BLE
is one-tenth lower than that of IP over WiFi; the performance depends on
the distance between devices and the message size.
Since this evaluation is limited, a comprehensive evaluation will be shown
in a future publication.

Besides, we have observed that the implementation on the Linux is not
mature enough.
Although our MQTT benchmark generates a none realistic workload, we found a
serious issue as described below.
Our MQTT benchmark have often failed, and any packets do not go through
the network after that until rebooting machines.  This issue was not observed
with wired and wireless Ethernet.
To pursue the cause, we have updated the Linux kernel from the version 3.18 to
4.0.  However, the situation does not change.  A stable implementation of
6LoWPAN over BLE is a future work.

Privacy is another big concern for IoT services.  We plan to develop MQTT
services using homomorphic encryption that allows computations to be carried
out on encrypted user data.  Such technology can extend the range of
application of IoT and 6LoWPAN.

\end{document}